\documentclass[seceq]{ptptex}

\usepackage{graphicx}
\usepackage{wrapft}

\title{Could Dense Quark Matter be a Source of Super High Energy Cosmic Rays?%        %You can use \\ for explicit line-break.
}

\author{Mais Suleymanov%       %Use \scshape for the family name.
}

\inst{  COMSATS  Institute of Information Technology , Islamabad, Pakistan %     %Affiliation, neglected when [addenda] or [errata].

}

\abst{ We propose that the dense quark matter could be a source of
the high-energy secondary hadrons.  These particles can be created
from hadronization of the parton(s), which possess the energy of
grouped partons from coherent interactions as a result of their
collective behavior in high dense medium. The medium
might be formed in the centre of some
massive stars, and it could be a source of the super high-energy
cosmic rays. In this work we consider some experimental results
as an evidence on collective phenomenon, that can lead to coherent
interactions in high dense medium and production of the high-energy
secondary hadrons.
}

\begin{document}

\maketitle

\section{Introduction}
%Start your paper from here.

%The talk focuses on one of the possible sources of the high energy
%cosmic particles and proposes the dense (and/or hot)  quark  matter
%as a possible source of the super high energy secondary hadrons  -
%super high energy cosmic rays.
Cosmic rays can provide us an important information on appearance
and evolution of the Universe. Since super high energy particle
beams (greater than $10^{17-18}~ eV$) are not available in
ground-based laboratories, super high energy cosmic rays are the
only resource to study interactions of the particles in this energy
domain. The source of super high energy cosmic are still
unknown~\cite{rf:1}.
%, moreover, we don't even know whether their
%origin is galactic or extragalactic.
%The flux of cosmic rays with energy up to  $~10^{10}~eV$ is mainly
%attributed to solar cosmic rays, intermediate energies (up to
%$~10^{15}~eV$) to galactic cosmic rays, and highest energies
%(greater than $10^{15}~eV$) to extragalactic cosmic rays.
The electromagnetic fields generated by some massive stars are
considered as plausible sources for the super high energy cosmic
rays~\cite{rf:2} , however, some theoretical predictions show that
these fields could be too weak to accelerate particles to energies
of order $10^{15}~eV$ . In conclusion, we see that it is necessary
to look for a new source mechanism of super high energy cosmic rays,
that doesn't involve acceleration.

\section{Conditions}

The dense quark matter can be a source of the super high energy
particles under following 3 conditions satisfied simultaneously:

1. Dense and/or hot quark matter with density   $\rho >> \rho_0$,
and/or with temperature $T >> T_0$ ( $\rho_0$ and $T_0$ are the
values of the density and the temperature of the normal nuclear
matter);

2. Collective behavior of partons in the medium and formation of
coherent parton group;

3. Coherent interaction in the system.

In these conditions the maximum energy of the produced partons are
limited only by the values of the total energy of the system, where
the values of energy will depend on the parameters of the system.

\section{What did we have until now?}

%Let us consider each of the conditions trying to understand what we
%had in the past and what we expect from Heavy Ion Collisions
%experiments at ultrarelativistic energies.

\subsection{First Condition - Dense and/or Hot Quark  Matter}

It is widely discussed that the dense and/or hot quark  matter can
be formed in the center of some massive  stars , for example as a
result of supernova explosion, and could lead to the  neutron stars
formation~\cite{rf:3}.

\subsection{Second Condition - Collective Behaviour}

%The experimental results on ultrarelativistic heavy ion collisions
%have shown the collective behavior for the partons in hot and dense
%matter.
%The $JINR$ Cumulative effect, $CERN~EMC$ effect  at
%relativistic hadron-nuclear and nuclear-nuclear interactions  can be
%considered as an experimental evidence on nucleon collective
%phenomenon in the medium.

\subsubsection{$JINR$ Cumulative Effect}

At relativistic energies we had the first signal on collective
behavior - $JINR$ Cumulative effect. It lead to the notion of
production of particles with energies beyond the kinematic limit of
free nucleon collisions~\cite{rf:4}. The effect was deeply discussed
in the paper~\cite{rf:5}  and below you can see some ideas from it.
Few interesting points:

- observation of the pions with energies $\backsimeq 8~GeV$ in $D+A$
reactions at $5~A~GeV$;

- in the $B+A\to C+X$   reactions the particles C were produced with
$x~>~1$.
%(see Fig.1)
The values of the  $x$ can be defined as $x={u\over s}\backsimeq
{{(\varepsilon - p cos \theta)}\over m }$ , here $u$ and $s$ are the
Mandelstam invariants, $m ,\varepsilon , p$ and $\theta$ are the
mass of nucleon, the total energies, the 3 momentum and the emission
angle for the $C$ particles  respectively, in the lab frame. For
free nucleon collisions the values of $x$ must be limited by $1$.
But as we can see from Fig.1~\cite{rf:6} for hadron-nuclear
interactions at $JINR$ energies particles were emitted with  $x>1$.

%%\begin{figure}[bth]
%%\parbox{\halftext}{% %\def\halftext{.471\textwidth}
%%%\figurebox{4cm}{2cm}
%%\includegraphics[width=4 cm,height=4 cm]
%%{figj1.eps} \caption{The first figure on the left.}} \hfill
%%\parbox{\halftext}
%%{
%%%\figurebox{4cm}{2cm}
%%\includegraphics[width=5 cm,height=4 cm]
%%{figj2.eps}\caption{The second figure on the right.}}
%%\end{figure}

\begin{figure}[bth]
\parbox{\halftext}{% %\def\halftext{.471\textwidth}
%\figurebox{4cm}{2cm}
\includegraphics[width=4.25 cm,height=3.0 cm]
{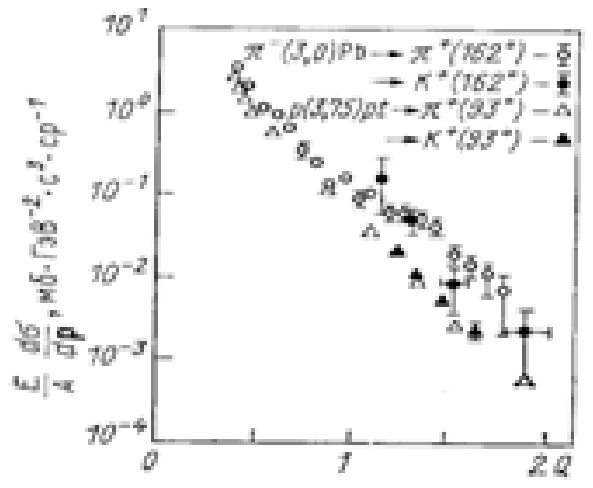} \caption{
%The first figure on the left.
}} \hfill
\parbox{\halftext}
{
%\figurebox{4cm}{2cm}
\includegraphics[width=4.25 cm,height=2.75 cm]
{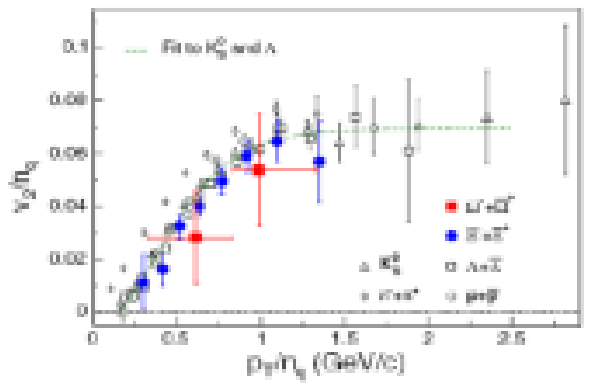}\caption{
%Number of quark ($n_q$) scaled $v_2$ as a
%function of scaled $p_T$ . All data are from $200~GeV~~Au+Au$
%minimum bias collisions.The dot-dashed-line is the scaled result of
%the fit to $K^0_S$ and $\Lambda$.
}}
\end{figure}

The JINR cumulative effect has very peculiar properties, some of
them are listed below:

1.  It  has been observed for photon-nuclear; lepton-nuclear;
hadron-nuclear and nuclear-nuclear interactions.

2. The strong $A$-dependences were indicated  for the invariant
inclusive cross sections of the cumulative particles(
$f(p)={\varepsilon {d\sigma\over d\bf p}}$ )

3.  The inverse of the slope for  $\varepsilon{d\sigma\over d\bf p}$
behavior  as a function of $x$ has a universal value  $<x>\backsimeq
0.16$

The theoretical interpretation of the effect proposed that it is a
result of nucleon collective phenomena and the cumulative particles
could be produced from the system of collected  nucleons- coherent
groups of nucleons. The latter could be formed as a result of
fluctuations of nuclear density ~\cite{rf:7} , the interaction of
the projectile with target nucleons
%(the possibility has a very
%small probability because the cumulative particles have been
%observed for photon and lepton interactions too)
, and    nucleon
percolation~\cite{rf:8}. So we can say that $JINR$ cumulative effect
can be considered as a phenomenon with nucleons collective behavior
and coherent interactions.

\subsubsection{$CERN~EMC$ Effect}

%The $JINR$ cumulative effect has been confirmed by the $CERN~EMC$
%effect~\cite{[9]}.

European Muon Collaboration ($EMC$) investigated the muon deep
inelastic scattering on iron and deuterium~\cite{rf:9}. They found
the big disagreement between experimental result and theoretical
expectations. The experimental result shown that the $F_2$ and hence
the quark and gluon distributions of a nucleon bound in a nucleus
differ from those of a free nucleon. None of the popular models
suggested to explain the $EMC$ effect seem satisfactory and present
a new point of view on the effect as a simple relativistic
phenomenon~\cite{rf:10}. The effect can also be considered as a
result of nucleon collective phenomena.

\subsubsection{ Azimuthal Anisotropy at $RHIC$ and $LHC$}
Azimuthal anisotropy observed experimentally at $RHIC$ and  $LHC$
shows a collective behavior, which is likely to be formed at an
early, parton, stage of the space-time evolution of the produced hot
and dense matter~\cite{rf:11}\tocite{rf:11dd}. The anisotropy
indicates that matter under extreme conditions behaves as a nearly
ideal liquid rather than an ideal gas of quarks and gluons. Scaling
behavior of $v_2$ vs $p_T$ ~\cite{rf:11d}\tocite{rf:11dd} gives a
possibility to assume that the collective behavior of the partons
defines the dynamics of the expansion in the longitudinal plane
namely (see Fig.2)

%%\begin{wrapfigure}{r}{4.0 cm} % r: RIGHT, 6.6cm: WIDTH
%%\figurebox{60mm}{3cm}
%%\includegraphics[width=4 cm,height=3 cm] {figj3.eps}
%%\caption{Number of quark ($n_q$) scaled $v_2$ as a function of
%%scaled $p_T$ . All data are from $200~GeV~~Au+Au$ minimum bias
%%collisions.
%% The dot-dashed-line is the scaled result of the fit to $K^0_S$ and $\Lambda$ . } \label{fig:2}
%%\end{wrapfigure}

The first measurement of elliptic  flow of charged particles in
$Pb-Pb$  collisions at  the center of mass  energy per nucleon pair
$\sqrt{s_{NN}} = 2.76~A~GeV$~\cite{rf:13} , with the $ALICE$
detector, demonstrated that the $v_2(p_t)$ does not change  within
uncertainties from the  $\sqrt{s_{NN}} = 200 GeV$ to $2.76~TeV$ .
$ALICE~LHC$ data demonstrated that values of the $v_2$ increase with
energy.

\subsection{Third Condition - Coherent Interactions}

%As we have mentioned above the $JINR$ cumulative effect could be
%explained as a result of nucleon collective phenomena and coherent
%interactions.
We have not had any experimental signal on coherent parton
interactions.  But we had the Coherent "Tube" Model
($CTM$)~\cite{rf:14}\tocite{rf:16}, which can give us even a clearer
explanation for the energetic (cumulative) particle production. Here
the interaction of a hadron with a target nucleus results from its
simultaneous collision with the tube of nucleons of cross section
$\sigma$ that lie along its path to the target nucleus. For the
interaction of projectile with momentum $p_{lab}$ the cumulative
square of the center-of-mass energy is $s_i\backsimeq 2imp_{lab}$
($i$ is a number of nucleons , $m$ - a nucleon mass). The
paper~\cite{rf:15}  quantitatively described unusually strong $A$
dependence (stronger than the commonly assumed $A$ or $A^{2/3}$) of
the cross section for $p + A\to J/\varPsi  + X$ reaction at incident
energies below $30~GeV$, using cumulative effects (via energy
rescaling). Ref.~\cite{rf:16} discusses the $CTM$ for high energy
nucleus-nucleus collisions. In this case two tubes are considered:
$i_1$ nucleons in incident tube and $i_2$ nucleons in the target
tube. The c.m. energy squared for this tube-tube collision is
approximately given by  $s_{i_1i_2}\backsimeq 2i_1i_2mp_{lab}$.

\section{Physical Picture}

So we could say that:

1.  In the high density (and/or high temperature) nuclear matter the
collective behaviour of partons could lead to formation of the
coherent groups of partons.
% ( "coherent tube") .

2.  As a result of the coherent interactions the parton(s) could be
produced with limited large values of $x\to 1$ and hardonize to
super high-energy particle(s), since in this case the resulting
energy will also depend on the parameters of the system, so the
energy of energetic particle can only be limited by values of the
total energy of the system.
%For example: in framework of $CTM$ the cumulative square of the
%center-of-mass energy $s$  will depend on a number $i$ of the
%partons, grouped in the tube(s), and  increase with $i$ . So if we
%consider a system with temperature around $150-170~MeV$ (hot matter)
%and density $7-10$ times greater than normal nucleus one (dense
%matter), then in this system the particle(s) with energy $
%10^{15}~eV$ could be produced as a result of coherent interactions
%of groped partons (or tube(s)) containing $ 10^7$ partons and more

3. The coherent interactions and formation of the super high energy
parton(s) can change the $x$ distribution of the partons.

4.  A medium with high density (and/ or high temperature) close to
the $QCD$ critical one could be a source of cosmic particles with
super high energies, and could be created in the center of some
massive star. The parton(s) with large values of $x$ or energy can
be formed in this system as a result of collective phenomenon and
coherent interactions,  hadronize and appear as super high-energy
cosmic particles. Their energies can only be limited by the values
of total energy of the grouped partons.

This physical picture assumes the existence of two strong
correlations in the hot and dense matter: between the partons with
limited small  values of $x\to 0$ ; between partons with limited
large values of  $x\to 1$ and limited small values of  $x\to 0$. It
means in hot and dense matter the $x$ distribution of partons can
vary and can get the structure with two additional maxima at $x\to
1$ and  $x\to 0$.

The observation of super high energy particles could be a signal on
the hot and dense states of strongly interacting matter , as well as
on the Quark Gluon Plasma formation.

%\section*{Summary}

%The physical picture suggests that the high density  (and/or high
%temperature) nuclear matter can contain strongly correlated
%multiparton formations and involve collective behaviour in the
%system, which  could lead to formation of the coherent groups of
%partons as a result of fluctuation of density  or percolation.  The
%system could be created in the center of some massive stars and can
%be a source of the super high energy cosmic particles. The
%experimental results on relativistic and ultrarelativistic
%hadron-nuclear and nuclear-nuclear interactions have pointed out the
%collective behavior of the partons. The new experimental results on
%coherent interactions of the partons are requested to test the
%physical picture. The latter assumes the existence of some
%correlations in dense (and/or hot) matter, which could be studied
%experimentally, and the centrality dependencies of the correlations
%could give essential information on the current physical picture.
%The observation of super high energy particles could be a signal on
%the hot and dense states of strongly interacting matter , as well as
%on the Quark Gluon Plasma formation.

%\section*{Acknowledgements}
%Author would like to thank HEC of Pakistan  for travel Grant for
%author's participation in the ISMD2011. Also thanks to A.M.
%Suleymanzade for help in preparing the paper.

%We would like to thank ...........

%\appendix
%\section{First Appendix} %Empty argument \section{} yields `Appendix'.
%
%\section{Second Appendix}

\end{document}